# Materials Science in Ancient Rome

Amelia Carolina Sparavigna
Dipartimento di Fisica,
Politecnico di Torino, Torino, Italy

Two books, the "De Architectura" by Vitruvius and the "Naturalis Historia" by Pliny the Elder, give us a portrait of the Materials Science, that is, the knowledge of materials, in Rome at the beginning of the Empire. Here, I am reporting some very attractive contents that we can find in these books. The reader will see the discussion proposed in fours case studies: concretes, coatings, amorphous materials and colloidal crystals, to describe them in modern words.

The American Chemical Society defines the Materials Science as an applied science studying the relationship between the structure and properties of materials. "Chemists who work in the field study how different combinations of molecules and materials result in different properties. They use this knowledge to synthesize new materials with special properties". The web page goes on remarking the fact that this science has a strong increasing impact on our society [1]. In fact, the Materials Science is an interdisciplinary field which, besides chemists, involves several researchers, mainly physicists and engineers, engaged to study the properties of matter, ranging from theoretical studies to technological applications, from the atomic scales to the macroscopic structures.

Since the interest of media is strongly attracted by nanoscience and nanotechnology, which are the most recent research fields of Materials Science, we have a tendency to forget that this science had always accompanied our life from quite ancient times. Nanoscience is its last development.

Science is coming from the Latin *scientia*, meaning "knowledge" [2], and then, if we want to find the Materials Science in the past, we need to investigate its "knowledge of materials". Let us then try to analyse the case of Ancient Rome. There are several studies on the roman technology (a quite impressive list is reported in Ref.3): my aim is to propose a research on available ancient texts of those parts which are devoted to the knowledge of materials. Of course, this is a huge task, needing help from scholars working on ancient Greek and Latin sources. Nevertheless, an initial investigation is possible on the books of Vitruvius and Pliny the Elder [4], giving quite interesting results.

Marcus Vitruvius Pollio lived in the 1st century BC. He was a Roman architect and engineer, best known as the author of the "De Architectura" [5]. Gaius Plinius Secundus, known as Pliny the Elder, died on 25 August, 79 AD during the eruption of Mount Vesusius. He was a Roman naturalist as well as naval commander of the Roman Empire. He wrote an encyclopaedic work, the "Naturalis Historia" [6]. These two books give us a portrait of knowledge in Rome, at the beginning of the Empire, and deserve a longer study. For instance, the "Naturalis Historia" contains several detailed descriptions of metals, pigments and precious stones. Here, I am reporting some contents that we can find in these books, which attracted my personal interest. In the following, the reader will find the discussion organized in some case studies on concretes, coatings, amorphous materials and colloidal crystals, to define them in modern words.

**1. From dust to stone**
Let us start to discuss the building materials, in particular, concretes.
Vitruvius devoted a chapter of his book to the pozzolana [5]. He is telling the following. "There is a species of sand which, naturally, possesses extraordinary qualities. It is found about Baiae and the territory in the neighbourhood of Mount Vesuvius; if mixed with lime and rubble, *it hardens as well under water as in ordinary buildings*. This seems to arise from the hotness of the earth under these mountains and the abundance of springs under their bases". Vitruvius is talking about pozzolana which is derived from unconsolidated pyroclastic materials erupted by the Campi Flegrei, a volcanic field, near Vesuvius. Vitruvius continues telling the "it seems certain that the moisture is extracted from the sand-stone and earth in their neighbourhood, by the strength of the fire, as from *limestone in a kiln*. Dissimilar and unequal actions thus concentrated towards the same end, the great want of moisture quickly supplied by water binds and strongly cements them, and also imparts a rapid solidity". But he notes that in Tuscany, where there are several hot springs too, "we do not there find a powder, which, for the same reason, would *harden under water* … All lands do not possess similar qualities; nor is stone universally found. Some lands are earthy, others gravelly, others gritty, others sandy: in short, the quality of land, in different parts of the earth, varies as much as even the climate itself."

What was written by Vitruvius is the precise description of, probably, the first "hydraulic cement". This cement hardens because of hydration, in reactions that occur independently of the mixture's water content. This cement is hence hardening underwater or exposed to wet weather. The results obtained when the anhydrous cement powder is mixed with water are not water-soluble [7].

Pozzolana is one of the components of the roman hydraulic concrete. The name pozzolana is derived from the "pulvis", that is dust of Puteoli and it is coming from pyroclastic materials. It was during the third century BC, that this hydraulic concrete was discovered by mixing pozzolanic materials with lime produced by heating limestone [8]. And in fact the harbour works at Puteoli were constructed (c.199 BC) with pozzolana cement. As reported in [8], the pozzolans are used as additives in modern portland cements too, improving the mechanical strength and providing resistance to weathering. It is necessary to note that deposits of pozzolans are not restricted to the Campi Flegrei or Vesuvius [8].

Pliny in his Natural History describes cements too. First of all, he starts to discuss the Opus signinum [9] and its manufacture, as follows. "What is there that human industry will not devise? Even broken pottery has been utilized; it being found that, beaten to powder, and tempered with lime, it becomes more solid and durable than other substances of a similar nature; forming the cement known as the "Signine" composition, so extensively employed for even making the pavements of houses. But there are other resources also, which are derived immediately from the earth. Who, indeed, cannot but be surprised at finding the most inferior constituent parts of it, known as "dust" ("pulvis") only, on the hills about Puteoli, forming a barrier against the waves of the sea, *becoming changed into stone the moment of its immersion*, and increasing in hardness from day to day — more particularly when mixed with the cement of Cumae? There is an earth too, of a similar nature found in the districts about Cyzicus; but there, it is not a dust, but a solid earth, which is cut away in blocks of all sizes, and which, after being immersed in the sea, is taken out transformed into stone. The same thing may be seen also, it is said, in the vicinity of Cassandrea; and at Cnidos, there is a spring of fresh water which has the property of causing earth to petrify within the space of eight months. Between Oropus and Aulis, every portion of the land upon which the sea encroaches becomes transformed into solid rock". As we can see, Pliny is also reporting information on the loci where it is possible to find this component of hydraulic cement or where similar phenomena can be observed, in a certain mythological fashion.

He continues "And then, besides, have we not in Africa and in Spain walls of earth, known as "formaceon" walls? from the fact that they are moulded, rather than built, by enclosing earth within a frame of boards, constructed on either side. These walls [10] will last for centuries, are proof against rain, wind, and fire, and are superior in solidity to any cement. Even at this day, Spain still beholds watch-towers that were erected by Hannibal, and turrets of earth placed on the very summits of her mountains." He is probably referring to the *tapias* [10-12], created with a mixture of lime, sand and gravel. This lime–cemented material was used, during the 1520s, to build the first palace of Cortés in Mexico City [12].

**2. Quicksilver and amalgams**
In his book, besides the history of metals, Pliny reports a huge quantity of minerals and pigments, mixing the description of their properties with history and mythology. Vitruvius is not so detailed in his description: let me start this section with his discussion on vermilion. We will end with a coating technique.

The vermilion is an opaque red pigment, similar to scarlet [13]. As a natural mineral pigment, it's known as Cinnabar, in particular from cinnabar mined in China. The chemical structure of the pigment is HgS mercuric sulfide. Vitruvius writes: "I shall now speak of vermilion … the manner of procuring and preparing it is very curious. A clod of earth is selected, which, before it is manufactured into vermilion, is called Anthrax, wherein are veins resembling iron, but of a red colour, and having a red dust round them. When dug up, it is beaten with iron bars till a great number of drops of quicksilver exude from it; these are immediately collected by the excavators. The clods, when collected in the laboratory, on account of their great dampness, are thrown into a furnace to dry; and the fumes that rise from them through the action of the fire fall condensed on the floor of the furnace, and are found to be quicksilver. But as, from the smallness of the drops which thus remain, they cannot be gathered up, they are swept into a vessel of water, in which they run together and re-unite. ... If quicksilver be placed in a vessel, and a stone of a hundred pounds weight be placed on it, it will swim at the top, and will, notwithstanding its weight, be incapable of pressing the liquid so as to break or separate it. If this be taken out, and only a single scruple of gold be put in, that will not swim, but immediately descend to the bottom. This is a proof that the *gravity of a body does not depend on its weight, but on its nature*."

Quite interesting this discussion, linked to the density of materials. Iron, quicksilver and gold have in fact increasing densities. We have not to be surprised: in the chapter on the method of detecting silver when mixed with gold, Vitruvius is reporting the tales on Archimedes discovering his principle during a bath. King Hiero was puzzled whether a crown was of pure gold or made of an alloy and he asked Archimedes to solve the problem. Archimedes found how to compare the densities of gold and silver, discovering a fraud.

Let us continue with the words of Vitruvius on quicksilver. "Quicksilver is used for many purposes; without it, neither silver nor brass can be properly gilt. When gold is embroidered on a garment which is worn out and no longer fit for use, the cloth is burnt over the fire in earthen pots; the ashes are thrown into water, and quicksilver added to them: this collects all the particles of gold, and unites with them. The water is then poured off, and the residuum placed in a cloth: which, when squeezed with the hands, suffers the liquid quicksilver to pass through the pores of the cloth, but retains the gold in a mass within it."

And now, let us see what Pliny is writing. "There is a mineral also found in these veins of silver, which yields a humour that is always liquid, and is known as "quicksilver." It acts as a poison upon everything, and pierces vessels even, making its way through them by the agency of its malignant properties (capillarity?). All substances float upon the surface of quicksilver, with the exception of gold, this being the only substance that it attracts to itself."  Here we have a new information on the knowledge of materials at ancient times: gold was the material with the highest specific gravity because platinum was unknown in Rome [6,14].

Quicksilver "it is such an excellent refiner of gold; for, on being briskly shaken in an earthen vessel with gold, it rejects all the impurities that are mixed with it. When once it has thus expelled these superfluities, there is nothing to do but to separate it from the gold: to effect which, it is poured out upon skins that have been well tawed, and so, exuding through them like a sort of perspiration, it leaves the gold in a state of purity behind."

As written in a footnote in [6], "the first use of quicksilver is commonly reckoned a Spanish invention, discovered about the middle of the sixteenth century; but it appears from Pliny (and let us add, from Vitruvius too), that the ancients were acquainted with amalgam and its use, not only for separating gold and silver from earthy particles, but also for gilding". And in fact, Pliny tells that "when copper has to be gilded, a coat of quicksilver is laid beneath the gold leaf, which it retains in its place with the greatest tenacity: in cases, however, where the leaf is single, or very thin, the presence of the quicksilver is detected by the paleness of the colour." We have then seen an ancient technique, for coating metals: in my opinion, a careful reading of Pliny's book can discover several other coating techniques.

**3. Glass and fire**
Here the amorphous material of Rome, the glasses Archaeological studies tell that probably the first man-made glass was created in coastal north Syria, Mesopotamia or Old Kingdom Egypt [15]. According to Pliny, Phoenician traders were the first to make the serendipitous discovery of the glass manufacturing technique. The episode is reported in the following manner. Pliny tells that in Syria there is a river, the Belus, which provides the pure sands good for produce glasses. The sand of this river was disclosed at the reflux of the tides. A ship, laden with nitre (a mineral alkali, [16]) was moored at its mouths. The Phoenician merchants, while cooking upon the sea-shore, "finding no stones at hand for supporting their cauldrons, employed for the purpose some lumps of nitre which they had taken from the vessel. Upon its being subjected to the action of the fire, in combination with the sand of the sea-shore, they beheld transparent streams flowing forth of a liquid hitherto unknown: this, it is said, was the origin of glass." As told in [15], this episode reflects the fact that probably, in the glass production, the white silica sand from this area was used because of its low impurity levels.

During the 1st century BC, glass blowing was discovered in the Syro-Palestinian area and the glass vessels became less expensive. The conquest of Judea by the Romans in 63 BC, prepared the growth of the use of glass products and their spread in the Roman world [15]. In the same reference, it is reported that the discovery of clear glass, through the use of manganese dioxide in the mixture, was made by the Jewish glass blowers in Alexandria, in the first century CE. The item is continuing telling that cast glass windows, albeit with poor optical qualities, began to appear in the most important buildings in Rome, Herculaneum and Pompeii. Let us remember that these two cities were buried by the Vesuvius volcanic eruption in AD 79. During this eruption Pliny died.

Pliny is telling in his book the following. "In process of time, as human industry is ingenious in discovering, it was not content with the combination of nitre, but magnet-stone (that is the Magnes lapis, Manganese, [16]) began to be added as well; from the impression that it attracts liquefied glass as well as iron. In a similar manner, too, brilliant stones of various descriptions came to be added in the melting, and, at last, shells and fossil sand." Assuming that Beckmann was right in considering that the name "Magnes lapis" comprehended manganese too [17], it means that the use of manganese dioxide in glass manufactures is surely antecedent 79 AD. The fact that Vitruvius does not discuss glasses seems to confirm that clear glasses were not available at his times.

Pliny continues in his description. "In fusing it, light and dry wood is used fur fuel, Cyprian copper and nitre being added to the melting, nitre of Ophir more particularly. It is melted, like copper, in contiguous furnaces, and a swarthy mass of an unctuous appearance is the result. ... This mass is again subjected to fusion in the furnace, for the purpose of colouring it; after which, the glass is either blown into various forms, turned in a lathe, or engraved like silver. Sidon was formerly famous for its glass-houses, for it was this place that first invented mirrors. Such was the ancient method of making glass: but, at the present day, there is found a very white sand for the purpose, at the mouth of the river Volturnus, in Italy. It spreads over an extent of six miles, upon the sea-shore that lies between Cumae and Liternum, and is prepared for use by pounding it with a pestle and mortar; which done, it is mixed with three parts of nitre, either by weight or measure, and, when fused, is transferred to another furnace. Here it forms a mass of what is called "hammonitrum"; which is again submitted to fusion, and becomes a mass of pure, white, glass.

At his time, according to Pliny, "the highest value is set upon glass that is entirely colourless and transparent, as nearly as possible resembling crystal, in fact. For drinking-vessels, glass has quite superseded the use of silver and gold (see Fig.1); but it is unable to stand heat unless a cold liquid is poured in first. And yet, we find that globular glass vessels, filled with water, when brought in contact with the rays of the sun become heated to such a degree as to cause articles of clothing to ignite." From this part of the text, it seems that he had not well understood that, in such case, the glass vessel acts as convex burning-glass. But this phenomenon was known. And probably glasses with water were used as magnifying glass.

Pliny is continuing his book, discussing the fire. "Submit to its action some sandy soil, and in one place it will yield glass, in another silver, in another minium, and in others, again, lead and its several varieties, pigments, and numerous medicaments. It is through the agency of fire that stones are melted into copper; by fire that iron is produced, and subdued to our purposes; by fire that gold is purified; by fire, too, that the stone is calcined, which is to hold together the walls of our houses. Some materials, again, are all the better for being repeatedly submitted to the action of fire; and the same substance will yield one product at the first fusion, another at the second, and another at the third. Charcoal, when it has passed through fire and has been quenched, only begins to assume its active properties; and, when it might be supposed to have been reduced to annihilation, it is then that it has its greatest energies." Is this the description of the production of activated charcoal, a form of carbon? I suppose so. Heating the charcoal is a process to make it extremely porous. Having a very large surface, the activated charcoal has enhance adsorption properties [18,19].

**4. Play of colour of opals**

Of course, several descriptions of precious stones are included in the "Natural History". Looking at their list, I found immediately some stones that attracted my interest, the opals. I asked myself if, in the Pliny's description, it was possible to find any mark of the modern vision of colloidal crystals. Pliny is telling "Of all precious stones, it is opal that presents the greatest difficulties of description, it displaying at once the piercing fire of carbuncle, the purple brilliancy of amethyst, and the sea-green of emerald, the whole blended together and refulgent with a brightness that is quite incredible. Some authors have compared the effect of its refulgence to that of the colour known as Armenian pigment, while others speak of it as resembling the flame of burning sulphur, or of flame fed with oil. ... There is no stone that is imitated by fraudulent dealers with more exactness than this, in glass, the only mode of detecting the imposition being by the light of the sun. For when a false opal is held between the finger and thumb, and exposed to the rays of that luminary, it presents but one and the same transparent colour throughout, limited to the body of the stone: whereas the genuine opal offers various refulgent tints in succession, and reflects now one hue and now another, as it sheds its luminous brilliancy upon the fingers."

For Pliny, the difficulty in describing this gemstone is due to the changing pattern of its colours. Gemmologists use the term "play of colour" to describe this effect, which is directly correlated by them to the value of the stone. Besides the play of colours, another part of the Pliny's description is quite attractive. It is the discussion on how to test the stone against frauds. At Pliny's times, as he is telling, the false opal was created from glass, probably with addition of some metal particles. Glass is an amorphous material: when we observe a piece of it by means of a light normally incident, without dispersion then, the object is transmitting just a colour. The behaviour of opal is different, and Pliny is remarkably describing: it has "refulgent tints in succession", reflecting different hues and shedding its colours upon the fingers.

In opal, the play of colour is caused by tiny sphere formations of silicon which make up the structure of opal [20,21]. The silica spheres have diameters ranging from 150 to 300 nm, arranged in a hexagonal or cubic close-packed lattice; they form colloidal crystals, quite attractive structures to create photonic crystals [21]. It is the regularity of the sizes and the packing of these spheres that determines the quality of precious opal (the arrangement of the spheres is shown in [22]).

The spheres arranged in a lattice structure create a diffraction grating for light [23]. The diffraction is controlled by the size of spheres and distances in the lattice. If the distance between the planes of the regularly packed spheres is approximately half the wavelength of a component of light, the light of that wavelength may be subject to diffraction. The spacing between the planes and their orientation with respect to the incident light determines the colours observed. The process can be described by Bragg's Law of diffraction. To imagine what happens, let us observe the reflecting surface of a compact disc: we see a play of colour, that is, a coloured appearance in succession from its surface. It is a diffraction of light again: the tracks of the compact disc act as a diffraction grating, separating the colours of white light. As told in Ref.23, the nominal track separation on a CD is 1.6 micrometers, giving about 625 tracks per millimeter. This corresponds to the range of laboratory diffraction gratings.

In the case of opals, as Pliny is telling, if the stone is thin enough, we see the diffraction patterns. Since the diffracted light cannot pass through large thicknesses of the opal, to display the play of colour the stone is often cut quite thin. This fact has given rise to an unusual method of preparing the stone as a gem: a thin layer of opal is backed by a swart mineral such as ironstone, basalt, or obsidian. This darker backing emphasizes the play of colour.

In this paper I have discussed just fours case studies, that I defined in modern words as: concretes, coatings, amorphous materials and colloidal crystals. I am confident that further readings of these books can provide further information on the knowledge of material in ancient Rome.

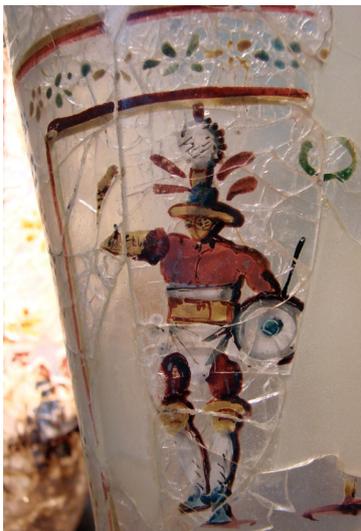

Fig.1 A beautiful example of Roman glass found at Bagram, Afghanistan (Gladiateur, verre émaillé, 1er siècle après JC, trésor de Begram, Musée Guimet, Paris, Courtesy: Vassil, http://en.wikipedia.org/wiki/File:Gladiateur_Begram_Guimet_18117.jpg).